\documentclass{article}
\usepackage{bbm}
\usepackage{mathrsfs}
\usepackage{fullpage,mathtools}
\usepackage{amsfonts}
\usepackage{graphicx}
\usepackage{amsmath}
\usepackage{amssymb}
\usepackage{dsfont}
\usepackage{float}
\usepackage{natbib}
\usepackage{bm}
\usepackage{subcaption} 
\usepackage{rotating,color}
\usepackage{xcolor}
\usepackage{booktabs}
\usepackage{amsmath}
\usepackage{amssymb}
\usepackage{amsfonts}
\usepackage{multirow}
\usepackage{amsthm}
\usepackage{authblk}
\newcommand{\R}{\mathbb{R}}
\newcommand{\E}{\mathbb{E}}

\newcommand{\Var}{\mathrm{Var}}
\newcommand{\Cov}{\mathrm{Cov}}

\newcommand{\tr}{\mathrm{tr}}

\newcommand{\cd}{\xrightarrow{d}}

\theoremstyle{plain}
\newtheorem{theorem}{Theorem}

\theoremstyle{definition}

\theoremstyle{remark}

\title{Rank-based Maxsum test for high dimensional regression coefficient}
\author{
Ping Zhao$^{a}$,
Liangliang Yuan$^{b,c}$\thanks{Corresponding author. Email: llyuan\_dlut@163.com}\\

\footnotesize
$^{a}$School of Mathematical Science, Tiangong University, China\\
$^{b}$School of Mathematics and Statistics, Nanjing University of Science and Technology, China\\
$^{c}$School of Statistics and Data Science, LEBPS, KLMDASR, and LPMC, Nankai University, China
}
\date{}

\begin{document}
\maketitle

\begin{abstract}
We study global inference for regression coefficients in high-dimensional linear models under potentially heavy-tailed errors.
While sum-type tests are powerful for dense alternatives and max-type tests excel for sparse alternatives, practical applications rarely reveal the sparsity level, and many existing procedures rely on light-tail assumptions.
Motivated by the Wilcoxon-score sum test of \citet{Feng2013} and the two Wilcoxon-score maximum tests of \citet{XuZhou2021}, we establish under $H_0$ the asymptotic independence between the rank-based sum statistic and each max statistic.
These joint limit results justify principled $p$-value aggregation, and we propose two adaptive \emph{rank-based maxsum} tests via the Cauchy combination method \citep{LiuXie2020CCT}.
The proposed procedures inherit robustness from rank-based construction and adaptivity from combining dense- and sparse-sensitive components.
Simulation studies confirm accurate size control and strong power across a wide range of error distributions and sparsity regimes.

{\it Keywords}: Cauchy combination test; High dimensional data; Maxsum test; Rank-based test; Wilcoxon score test.
\end{abstract}

\section{Introduction}\label{sec:intro}

High-dimensional linear regression has become a basic modeling tool across modern data-driven sciences, where the number of covariates can be comparable to or far larger than the sample size.
Typical examples include large-scale genetic association studies and genotype--phenotype mapping \citep{Chen2006,Ballard2010,Chiang2006}, high-throughput gene expression and pathway analysis \citep{Scheetz2006,Subramanian2005}, and data-rich problems in drug discovery and bioavailability prediction \citep{Horn2007,Redfern1999}.
In such applications, a fundamental inferential task is to test whether the covariates collectively contribute to the response, namely the global hypothesis $H_0:\bm\beta=\bm 0$ in the high-dimensional linear model.
Reliable global testing procedures are essential not only for scientific discovery but also for downstream procedures such as screening, model building, and resource allocation.

A natural starting point is to generalize the classical low-dimensional $F$-test to high dimensions.
\citet{Goeman2006} proposed a global test against high-dimensional alternatives (and developed asymptotic type-I error control in generalized linear models \citep{Goeman2011}), offering a principled extension of score-type testing ideas to settings where $p$ is large.
In linear regression with structured designs, \citet{Zhong2011} developed tests that can be viewed as high-dimensional analogues of the classical $F$-statistic.
However, a common limitation of such sum- or quadratic-form-type procedures is that their power is primarily driven by \emph{dense} alternatives, where many coordinates of $\bm\beta$ deviate from zero.
When the signal is \emph{sparse} (only a small subset of coefficients is nonzero), maximum-type procedures are often substantially more powerful.
This sparsity--density tradeoff has motivated the development of complementary max-type tests and, importantly, combination strategies that remain powerful without knowing the underlying sparsity level.
A recent representative line of work establishes asymptotic independence between suitably standardized sum-type and max-type statistics under $H_0$, which enables simple combination rules such as selecting the minimum of the component $p$-values \citep{FengJiangLiLiu2024}.

Most of the aforementioned high-dimensional regression tests are derived under light-tailed error assumptions (Gaussian or sub-Gaussian-like) and can exhibit degraded performance when errors are heavy-tailed or contaminated.
To address robustness, \citet{Feng2013} extended classical rank-based inference \citep{McKean1976,Hettmansperger1998} to high-dimensional regression by proposing a Wilcoxon-score-based global test.
This rank-based sum-type test is distribution-free under $H_0$ (under mild regularity) and shows strong empirical performance under heavy-tailed errors, especially for dense alternatives.
Complementing this, \citet{XuZhou2021} introduced two Wilcoxon-score-based maximum-type tests for high-dimensional regression coefficients, tailored to sparse alternatives: a marginal max statistic and a precision-transformed max statistic designed to accommodate predictor dependence.
Despite their complementary strengths, the sparsity level is unknown in practice, so a single procedure that is simultaneously robust to heavy tails and adaptive to sparsity is highly desirable.

This paper develops such adaptive and robust global tests by combining the rank-based sum-type statistic of \citet{Feng2013} with the Wilcoxon-score-based max-type statistics of \citet{XuZhou2021}.
Our key theoretical contribution is to prove, under $H_0$, the \emph{asymptotic independence} between the Wilcoxon sum-type statistic and each of the two Wilcoxon max-type statistics.
This joint limit theory justifies principled combination of the corresponding $p$-values.
Building on the Cauchy combination methodology \citep{LiuXie2020CCT}, we propose two \emph{rank-based maxsum} tests that fuse evidence from dense- and sparse-sensitive components.
The resulting procedures are (i) robust to heavy-tailed and non-Gaussian errors by construction, and (ii) adaptive to unknown sparsity through maxsum combination.
Extensive simulations further demonstrate that the proposed maxsum tests maintain stable size and deliver strong power across a wide range of error distributions and sparsity regimes, improving upon using either component alone.

The remainder of the paper is organized as follows.
Section~\ref{sec:main} states two asymptotic independence theorems under $H_0$ and presents the proposed Cauchy-combination maxsum tests.
Section~\ref{sec:sim} reports simulation results.
Technical proofs are collected in the Appendix.

\section{Rank-based Maxsum Tests}\label{sec:main}
We consider the high-dimensional linear regression model
\begin{equation}\label{eq:model}
Y_i=\bm X_i^\top \bm\beta+\varepsilon_i,  i=1,\ldots,n,
\end{equation}
where $\bm X_1,\ldots,\bm X_n$ are i.i.d.\ $p$-dimensional covariates and $\bm\beta\in\R^p$ is the regression coefficient vector.
The noise $\varepsilon_i$ is independent of $\bm X_i$.
Throughout, we assume the covariates are standardized so that $\E(\bm X_i)=\bm 0_p$ and $\Var(X_{ik})=1$ for $k=1,\ldots,p$.
Let $\Sigma=\Var(\bm X_i)$ and assume $\Sigma$ is positive definite to ensure identifiability.
Our goal is to test the global hypothesis
\begin{equation}\label{eq:hypo}
H_0:\bm\beta=\bm\beta_0 \quad \text{vs}\quad H_1:\bm\beta\neq \bm\beta_0,
\end{equation}
for a given $\bm\beta_0\in\R^p$; without loss of generality we take $\bm\beta_0=\bm 0_p$.

When $p<n$, the classical $F$-test provides a standard solution. 
In high dimensions, however, the $F$-test may suffer substantial power loss as $p$ approaches $n$ and becomes ill-defined once the relevant sample covariance matrices are singular (e.g., when $p\ge n$).
To circumvent the inversion of $\bm X^\top \bm X$ in the classical $F$-statistic, \citet{Goeman2006} introduced an empirical-Bayes-type global test based on
\begin{equation}\label{eq:eb_rewrite}
T_{EB}=\frac{\bm Y^\top \bm X \bm X^\top \bm Y}{\bm Y^\top \bm Y},
\end{equation}
which can be interpreted as a score test for a hyper-parameter in a prior placed on $\bm\beta$.
Along a related direction, \citet{Zhong2011} proposed a centered $U$-statistic-based procedure,
\[
Z_n=\frac{1}{4P_n^4}\sum^{*}
(\bm X_{i_1}-\bm X_{i_2})^\top(\bm X_{i_3}-\bm X_{i_4})
(Y_{i_1}-Y_{i_2})(Y_{i_3}-Y_{i_4}),
\]
where $\sum^{*}$ sums over mutually distinct indices and $P_n^4=n!/(n-4)!$, and established asymptotic normality under a diverging factor structure \citep{Bai1996}.
Both $T_{EB}$ and $Z_n$ are representative \emph{sum-type} (quadratic-form/U-statistic) tests, which are typically effective against \emph{dense} alternatives but can be less sensitive when the signal is \emph{sparse}.

To enhance power under sparse signals, \citet{FengJiangLiLiu2024} developed a \emph{maximum-type} statistic of the form
\begin{equation}\label{eq:tmax_rewrite}
T_{MAX}=\max_{1\le j\le p}\left(\frac{1}{\sqrt n}\,\mathbb X_j^\top \bm Y\right)^2\big/\hat\sigma^2,
\end{equation}
where $\mathbb X_j$ is the $j$th column of $\bm X$ and $\hat\sigma^2$ consistently estimates $\Var(\varepsilon_i)$.
Moreover, they proved asymptotic independence between a max-type component and a sum-type component (e.g., $T_{EB}$), which enables adaptive testing by combining information from both regimes via a minimum-$p$ strategy.

The above procedures are largely motivated by light-tailed noise settings and may lose robustness when $\varepsilon_i$ is heavy-tailed.
To address this issue, \citet{Feng2013} extended rank-based methodology to high-dimensional regression.
Let $R_i$ be the rank of $Y_i$ among $\{Y_1,\ldots,Y_n\}$ and define the Wilcoxon scores
\begin{equation}\label{eq:wilcoxon_rewrite}
e_i=\sqrt{12}\left(\frac{R_i}{n+1}-\frac12\right),  i=1,\ldots,n,
\end{equation}
with $\bm e=(e_1,\ldots,e_n)^\top$.
Under $H_0$ and continuity, $\{e_i\}$ are distribution-free with $\E(e_i)=0$ and $\Var(e_i)=1$.
They proposed a rank-score $U$-statistic
\begin{equation}\label{eq:wn_rewrite}
W_n=\frac{1}{n(n-1)}\sum_{1\le i<k\le n}\bm X_i^\top\bm X_k\,e_i e_k,
\end{equation}
and the standardized sum-type statistic
\begin{equation}\label{eq:trs_rewrite}
T_{RS}=\frac{nW_n}{\sqrt{2\widehat{\tr(\mathbf{\Sigma}^2)}}},
\end{equation}
where $\widehat{\tr(\mathbf{\Sigma}^2)}$ is an consistent estimator of $\tr(\mathbf{\Sigma}^2)$ \citep{Zhong2011,Feng2013}, i.e. 
$$
\widehat{\operatorname{tr}\left(\mathbf{\Sigma}^2\right)}=\frac{1}{2 P_n^4} \sum^*\left(\bm{X}_{i_1}-\bm{X}_{i_2}\right)^T\left(\bm{X}_{i_3}-\bm{X}_{i_4}\right)\left(\bm{X}_{i_3}-\bm{X}_{i_2}\right)^T\left(\bm{X}_{i_1}-\bm{X}_{i_4}\right) .
$$
This rank-based sum statistic is robust to heavy tails and is particularly competitive for dense alternatives.

For sparse alternatives under potentially non-Gaussian errors, \citet{XuZhou2021} proposed two Wilcoxon-score-based max-type tests.
The first one is the marginal maximum statistic
\begin{equation}\label{eq:trm1_rewrite}
T_{RM1}=\max_{1\le j\le p} w_j^2,
\end{equation}
where $w_j$ is a Studentized rank-score correlation from regressing the Wilcoxon scores on the centered covariate $\mathbb X_j$,
\[
w_j=c_n\,\bm e^\top\!\left(\mathbb X_j-\mathbf 1_n\mathbf 1_n^\top \mathbb X_j/n\right)\Big/
\left\|\mathbb X_j-\mathbf 1_n\mathbf 1_n^\top \mathbb X_j/n\right\|, 
c_n=\sqrt{\frac{n+1}{n-1}}.
\]
To further improve power in the presence of dependence among covariates, they introduced a precision-adjusted version
\begin{equation}\label{eq:trm2_rewrite}
T_{RM2}=\max_{1\le j\le p} W_j^2,
\end{equation}
where $W_j$ is constructed from $\bm X\hat{\mathbf{\Omega}}=(\mathbb X_1^\natural,\ldots,\mathbb X_p^\natural)$ with an estimator of $\mathbf{\Omega}=\mathbf{\Sigma}^{-1}=(\omega_{kl})$ and
\[
W_j=c_n\,\bm e^\top\!\left(\mathbb X_j^\natural-\mathbf 1_n\mathbf 1_n^\top \mathbb X_j^\natural/n\right)\Big/\sqrt{n\hat\omega_{jj}}.
\]
In implementation, $\hat{\mathbf{\Omega}}$ can be estimated using high-dimensional covariance/precision matrix methodology; see, for example, \citet{Bickel2008,Cai2011}.

Since $T_{RS}$ is tailored to dense alternatives whereas $T_{RM1}$ and $T_{RM2}$ are designed for sparse alternatives, a data-adaptive strategy is needed when the sparsity level is unknown.
To this end, we establish under $H_0$ the asymptotic independence between the rank-based sum statistic $T_{RS}$ and each of the two rank-based max statistics $T_{RM1}$ and $T_{RM2}$, respectively.
These results enable principled $p$-value aggregation, and we will further develop two robust and sparsity-adaptive maxsum tests based on the Cauchy combination idea.

To establish the asymptotic independence result, we impose the following assumptions:
\begin{itemize}
\item[(C1)] The regressors follow an independent component model:
$
\bm X_i=\mathbf{\Sigma}^{1/2}\bm Z_i,  \bm Z_i=(Z_{i1},\ldots,Z_{ip})^\top,
$
where $Z_{i1},\ldots,Z_{ip}$ are i.i.d., and there exist positive constants $C$ and $K$ such that
$
\mathbb E\{\exp(Z_{ij}^2/C)\}\le K  \text{for all } i,j .
$

\item[(C2)] Let $\mathbf{\Sigma}=(\sigma_{kl})_{1\le k,l\le p}$ and $\mathbf{\Omega}=\mathbf{\Sigma}^{-1}=(\omega_{kl})_{1\le k,l\le p}$. The diagonal entries $\sigma_{kk}$ and $\omega_{ll}$ are uniformly bounded away from $0$ and $\infty$ for $1\le k,l\le p$. Moreover, for some $c_0<1$,
$
\max_{1\le k<l\le p}\frac{|\sigma_{kl}|}{\sqrt{\sigma_{kk}\sigma_{ll}}}\le c_0,
\max_{1\le k<l\le p}\frac{|\omega_{kl}|}{\sqrt{\omega_{kk}\omega_{ll}}}\le c_0,
$
and for some constant $c>0$,
$
\max_{1\le l\le p}\sum_{k=1}^p \sigma_{kl}^2\le c,
\max_{1\le l\le p}\sum_{k=1}^p \omega_{kl}^2\le c.
$
\item[(C3)] The error $\varepsilon$ admits a density $f(\cdot)$ whose derivatives are uniformly bounded, and $f$ has finite Fisher information:
$
\int f^{-1}(x)\{f'(x)\}^2\,dx<\infty .
$
\end{itemize}

Condition (C1) specifies the dependence structure and tail behavior of the regressors. In particular, the independent component representation together with the sub-exponential tail control is sufficient for the diverging factor-type conditions used in \citet{Feng2013}. Moreover, when combined with the covariance/precision sparsity in (C2), it also aligns with the type of regressor regularity required in \citet{XuZhou2021}.

Condition (C2) imposes a joint sparsity/weak-dependence requirement on both the covariance matrix $\mathbf{\Sigma}$ and the precision matrix $\mathbf{\Omega}$, ensuring that correlations and partial correlations are uniformly bounded away from $1$ and that each column has bounded $\ell_2$-energy. These constraints are standard for deriving non-degenerate limiting null distributions for max-type statistics in high dimensions. Under (C2), commonly used trace conditions in sum-type tests---such as $\tr(\mathbf{\Sigma}^4)=o\{\tr^2(\mathbf{\Sigma}^2)\}$ in \citet{Feng2013}---are also satisfied.

Condition (C3) is an error distribution assumption typical in rank-based inference (e.g., \citet{Hettmansperger1998,Feng2013}). It guarantees sufficient smoothness for the underlying score/influence-function expansions while only requiring finite Fisher information; notably, it does not require $\varepsilon$ to have a bounded variance, allowing for heavy-tailed errors.

\begin{theorem}\label{th1}
Assume conditions (C1)-(C3) hold. Under $H_0$, if $\log p=o(n^{1/5})$, we have
\begin{align}
P\left(T_{RS}\le x, T_{RM1}-2\log p+\log\log p\le y\right)\to \Phi(x)F(y)
\end{align}
where $\Phi(x)$ is the c.d.f of $N(0,1)$ and $F(y)=\exp(-\frac{1}{\sqrt{\pi}}e^{-y/2})$ is the c.d.f of Gumble distribution.
\end{theorem}

\begin{theorem}\label{th2}
    Assume conditions (C1)-(C3) hold. We assume that the estimator $\hat{\mathbf{\Omega}}=\left(\hat{\mathbf{\omega}}_{ij}\right)$ has at least a logarithmic rate of convergence
$
\|\hat{\mathbf{\Omega}}-\mathbf{\Omega}\|_{L_1}=o_p\left\{\frac{1}{\log (p)}\right\}, \quad \max _{1 \leq i \leq p}\left|\hat{\omega}_{ii}-\omega_{ii}\right|=o_p\left\{\frac{1}{\log (p)}\right\} .
$
Under $H_0$, if $\log p=o(n^{1/5})$, we have
\begin{align}
P\left(T_{RS}\le x, T_{RM2}-2\log p+\log\log p\le y\right)\to \Phi(x)F(y).
\end{align}
\end{theorem}
Based on Theorems~\ref{th1} and~\ref{th2}, we further construct two Cauchy combination tests
via the Cauchy combination principle of \citet{LiuXie2020CCT}. Let $p_{RS}$, $p_{RM1}$ and $p_{RM2}$
denote the $p$-values associated with the rank-sum type statistic $T_{RS}$ and the two max-type
rank statistics $T_{RM1}$ and $T_{RM2}$, respectively. We define
\begin{align}
T_{RC1}
&=
\frac{1}{2}\tan\!\left\{\Big(\tfrac{1}{2}-p_{RS}\Big)\pi\right\}
+\frac{1}{2}\tan\!\left\{\Big(\tfrac{1}{2}-p_{RM1}\Big)\pi\right\},
\label{eq:TRC1}\\
T_{RC2}
&=
\frac{1}{2}\tan\!\left\{\Big(\tfrac{1}{2}-p_{RS}\Big)\pi\right\}
+\frac{1}{2}\tan\!\left\{\Big(\tfrac{1}{2}-p_{RM2}\Big)\pi\right\}.
\label{eq:TRC2}
\end{align}

By Theorems~\ref{th1}--\ref{th2}, the component $p$-values are asymptotically valid under $H_0$,
and the corresponding sum-type and max-type statistics are asymptotically independent.
Therefore, applying the main result of \citet{LiuXie2020CCT}, both $T_{RC1}$ and $T_{RC2}$
converge in distribution to the standard Cauchy distribution under the null hypothesis, that is,
$$
T_{RC1}\cd \mathrm{Cauchy}(0,1),
\qquad
T_{RC2}\cd \mathrm{Cauchy}(0,1).
$$
Consequently, for a given significance level $\alpha\in(0,1)$, we reject $H_0$ if
$$
T_{RC1}>c_\alpha
\quad\text{or}\quad
T_{RC2}>c_\alpha,
$$
respectively, where $c_\alpha$ is the upper $\alpha$-quantile of the standard Cauchy distribution.
Equivalently, since $P(\mathrm{Cauchy}(0,1)>t)=\frac{1}{\pi}\arctan(t)^{\!\!-1}$, one may report the
combined $p$-value as
$$
p_{RC1}=\frac{1}{2}-\frac{1}{\pi}\arctan(T_{RC1}),
\qquad
p_{RC2}=\frac{1}{2}-\frac{1}{\pi}\arctan(T_{RC2}),
$$
and reject $H_0$ whenever $p_{RC1}<\alpha$ or $p_{RC2}<\alpha$.

The proposed Cauchy combinations inherit the strengths of both components: the rank-sum part
is powerful against dense and moderately sparse alternatives, while the max-type part is tailored
to very sparse signals. As a result, $T_{RC1}$ and $T_{RC2}$ provide adaptive power improvement
over any single test, while maintaining asymptotic level $\alpha$ under $H_0$ without requiring
resampling or explicit dependence estimation between the component statistics.

\section{Simulation}\label{sec:sim}

We compare the following eight procedures:
\begin{itemize}
\item[(i)] \textbf{Max}: the max-type test proposed by \citet{FengJiangLiLiu2024};
\item[(ii)] \textbf{EB}: the score-type test of \citet{Goeman2006};
\item[(iii)] \textbf{COM}: the adaptive combination test proposed by \citet{FengJiangLiLiu2024};
\item[(iv)] \textbf{RS}: the Wilcoxon rank-score sum-type test of \citet{Feng2013};
\item[(v)] \textbf{RM1}: the Wilcoxon-score max-type test $T_w$ of \citet{XuZhou2021}, with Gaussian multiplier approximation;
\item[(vi)] \textbf{RM2}: the Wilcoxon-score max-type test $T_W$ of \citet{XuZhou2021}, where the precision matrix $\Omega$ is estimated via banding with cross-validated tuning (implemented by \texttt{CVTuningCov});
\item[(vii)] \textbf{RC1}: the Cauchy combination of the $p$-values from \textbf{RS} and \textbf{RM1};
\item[(viii)] \textbf{RC2}: the Cauchy combination of the $p$-values from \textbf{RS} and \textbf{RM2}.
\end{itemize}
For the Gaussian multiplier approximations, we use $B=2000$ multipliers, matching the implementation in our code.

The covariate vectors $\bm X_i\in\mathbb R^p$ are generated from a mean-zero Gaussian distribution with an AR(1)-type dependence:
\[
\bm X_i \sim N(\mathbf 0,\Sigma),  \Sigma_{jk}=0.7^{|j-k|},\quad 1\le j,k\le p.
\]
We consider $(n,p)\in\{(100,200),(100,400),(200,200),(200,400)\}$. 
To examine robustness, we generate i.i.d.\ errors $\{\varepsilon_i\}_{i=1}^n$ from four distributions:
\begin{enumerate}
\item[(E1)] normal distribution: $\varepsilon_i\sim N(0,1)$;
\item[(E2)] standardized $t$ with 3 degrees of freedom: $\varepsilon_i=t_{3}/\sqrt{3}$ ;
\item[(E3)] standardized log-normal: $\varepsilon_i=\{ \exp(Z_i)-\exp(1/2)\}/\sqrt{\exp(1)\{\exp(1)-1\}}$, where $Z_i\sim N(0,1)$ ;
\item[(E4)] standardized Gaussian mixture: $\varepsilon_i \sim \{0.9N(0,1)+0.1N(0,100)\}/\sqrt{10.9}$.
\end{enumerate}

Under $H_0$, we set $\bm \beta=\mathbf 0$ so that $Y_i=\varepsilon_i$ and $Y$ is independent of $X$. 
For each configuration $(n,p)$ and each error distribution (E1)--(E4), we run $1000$ Monte Carlo replications and report the empirical size at the nominal level $\alpha=0.05$ (i.e., the rejection frequency).

Table~\ref{tab:size_8methods} reports the empirical sizes of all eight methods. Overall, the procedures exhibit satisfactory size control across the considered $(n,p)$ pairs and error distributions, including heavy-tailed and contaminated cases (E2)--(E4). 
The max-type test  {Max} is mildly conservative in several settings, while  {EB} and  {RM1} tend to be close to the nominal level. 
The rank-score sum test  {RS} shows slight size inflation in a few scenarios, and consequently the Cauchy combinations  {RC1}-- {RC2} can also be somewhat liberal when one component $p$-value departs from uniformity in finite samples. 
Nevertheless, the observed deviations are moderate, indicating that the proposed procedures provide reliable calibration for high-dimensional coefficient testing under the dependence structure considered here.

\begin{table}[!htbp]
\centering
\caption{Empirical size (\%) of eight methods.}
\label{tab:size_8methods}
\footnotesize
\setlength{\tabcolsep}{6pt}
\begin{tabular}{cccccccccc}
\toprule
$n$ & $p$ & Max & EB & COM & RS & RM1 & RM2 & RC1 & RC2 \\
\midrule
\multicolumn{10}{c}{E1}\\ \hline
100 & 200 & 3.3 & 5.0 & 3.1 & 5.8 & 4.5 & 4.1 & 6.1 & 5.8 \\
100 & 400 & 2.7 & 4.5 & 2.9 & 5.4 & 3.8 & 5.8 & 5.2 & 5.9 \\
200 & 200 & 2.1 & 5.6 & 4.1 & 7.0 & 3.6 & 5.3 & 6.7 & 6.0 \\
200 & 400 & 2.8 & 4.6 & 4.3 & 6.3 & 4.3 & 4.7 & 5.9 & 6.3 \\ \hline
\multicolumn{10}{c}{E2}\\ \hline
100 & 200 & 2.4 & 3.8 & 2.9 & 5.8 & 3.7 & 6.1 & 5.2 & 6.2 \\
100 & 400 & 2.6 & 5.1 & 3.4 & 6.5 & 4.5 & 5.5 & 6.8 & 6.8 \\
200 & 200 & 2.3 & 4.2 & 3.9 & 6.5 & 3.3 & 4.3 & 5.8 & 6.1 \\
200 & 400 & 3.6 & 6.6 & 4.8 & 7.8 & 4.7 & 4.8 & 6.9 & 6.7 \\ \hline
\multicolumn{10}{c}{E3}\\ \hline
100 & 200 & 3.1 & 6.6 & 4.3 & 5.8 & 4.3 & 4.2 & 6.1 & 5.8 \\
100 & 400 & 2.4 & 5.0 & 3.2 & 5.3 & 3.8 & 5.4 & 5.1 & 5.5 \\
200 & 200 & 2.4 & 6.4 & 4.4 & 6.9 & 3.4 & 5.3 & 6.6 & 6.9 \\
200 & 400 & 2.7 & 4.9 & 2.9 & 6.4 & 4.2 & 4.8 & 5.8 & 6.3 \\ \hline
\multicolumn{10}{c}{E4}\\ \hline
100 & 200 & 3.4 & 4.7 & 3.7 & 6.6 & 4.4 & 4.9 & 6.5 & 6.2 \\
100 & 400 & 3.8 & 4.9 & 3.8 & 5.3 & 4.1 & 5.8 & 5.7 & 6.3 \\
200 & 200 & 2.2 & 4.4 & 3.9 & 7.5 & 5.1 & 4.7 & 7.1 & 6.3 \\
200 & 400 & 3.2 & 4.4 & 4.1 & 6.1 & 4.6 & 5.1 & 5.2 & 5.6 \\
\bottomrule
\end{tabular}
\end{table}

We compare the power of the tests with $n=100, p=200$. Define $\boldsymbol{\beta}_b=\kappa \cdot\left(\beta_{q+1}, \cdots, \beta_p\right)^T$. Let $s$ denote the number of nonzero coefficients. For $s=1, \cdots, m$, we consider $\beta_j \sim N(0,1), q< j \leqslant q+s$ and $\beta_j=0, j>q+s$. The parameter $\kappa$ is chosen so that $\left\|\boldsymbol{\beta}_b\right\|^2=0.8$.  

\begin{figure}[!htbp]
\centering
\begin{subfigure}[t]{0.48\textwidth}
  \centering
  \includegraphics[width=\linewidth]{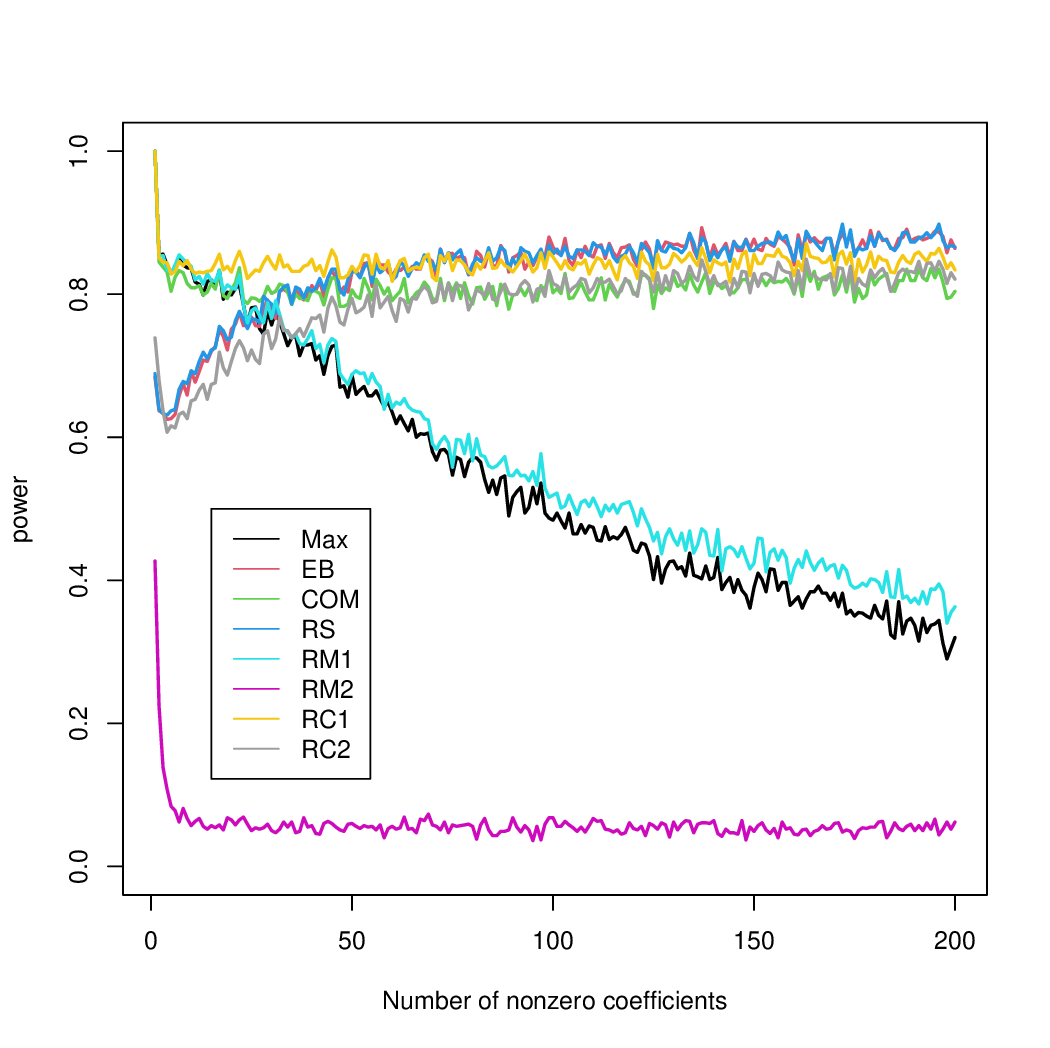}
  \caption{Normal distribution}
  \label{fig:E1}
\end{subfigure}\hfill
\begin{subfigure}[t]{0.48\textwidth}
  \centering
  \includegraphics[width=\linewidth]{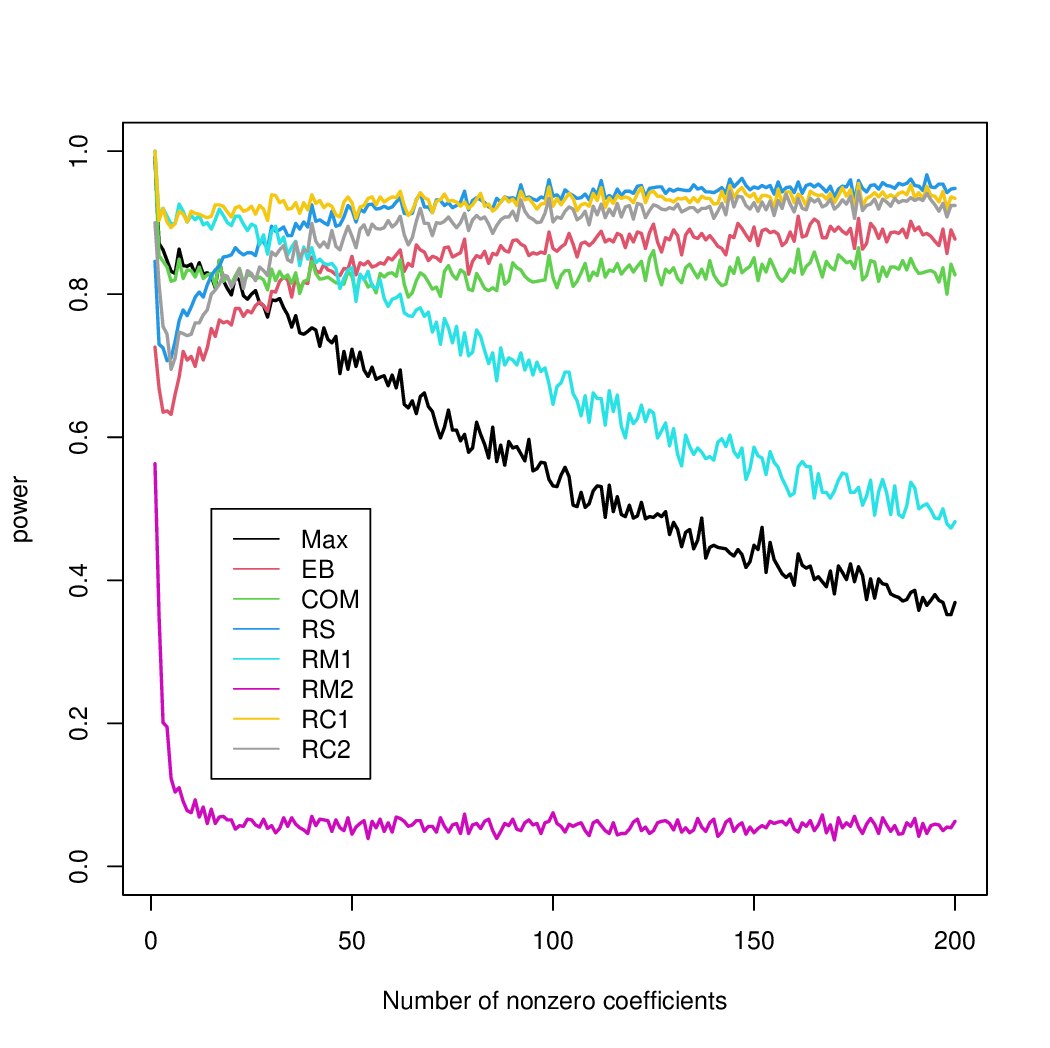}
  \caption{t(3) distribution}
  \label{fig:E2}
\end{subfigure}

\vspace{2mm}

\begin{subfigure}[t]{0.48\textwidth}
  \centering
  \includegraphics[width=\linewidth]{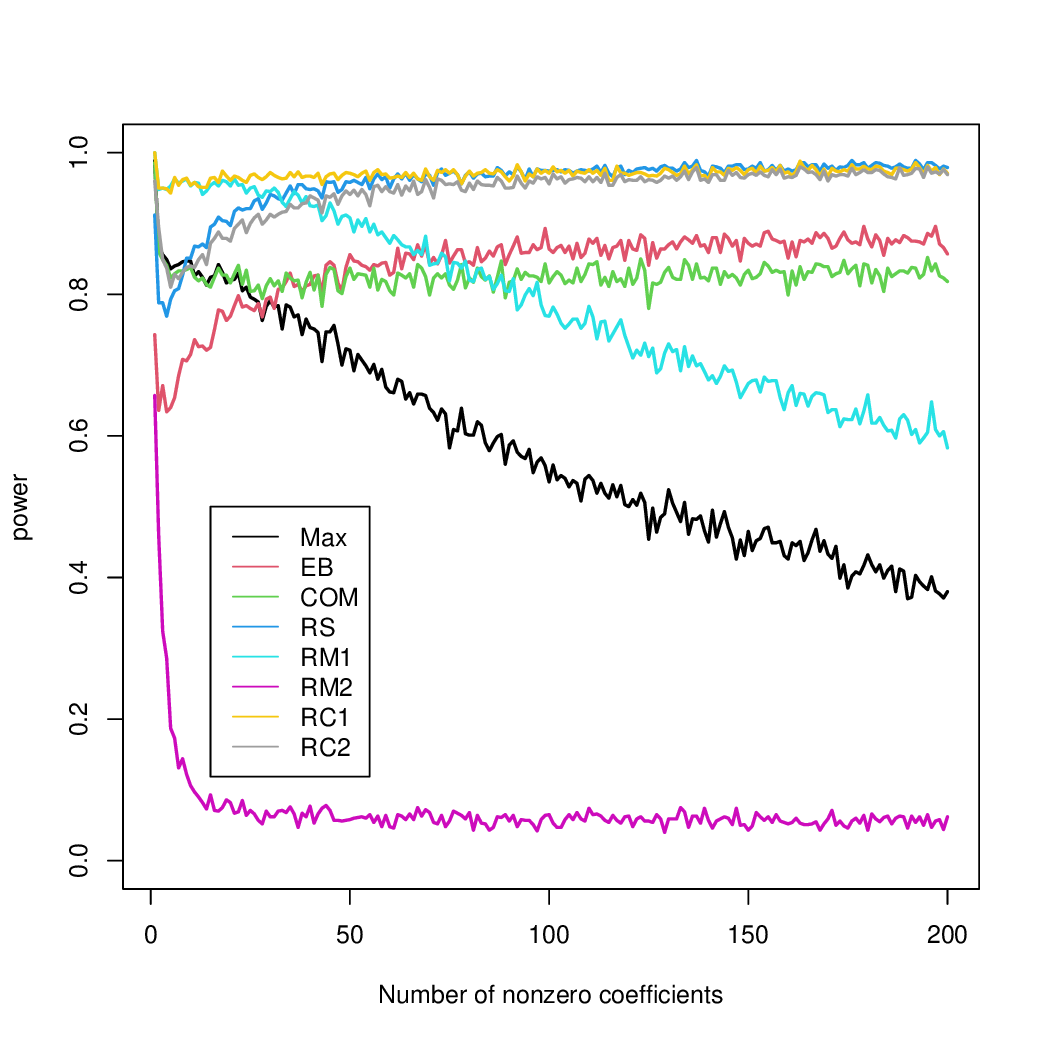}
  \caption{Lognormal distribution}
  \label{fig:E3}
\end{subfigure}\hfill
\begin{subfigure}[t]{0.48\textwidth}
  \centering
  \includegraphics[width=\linewidth]{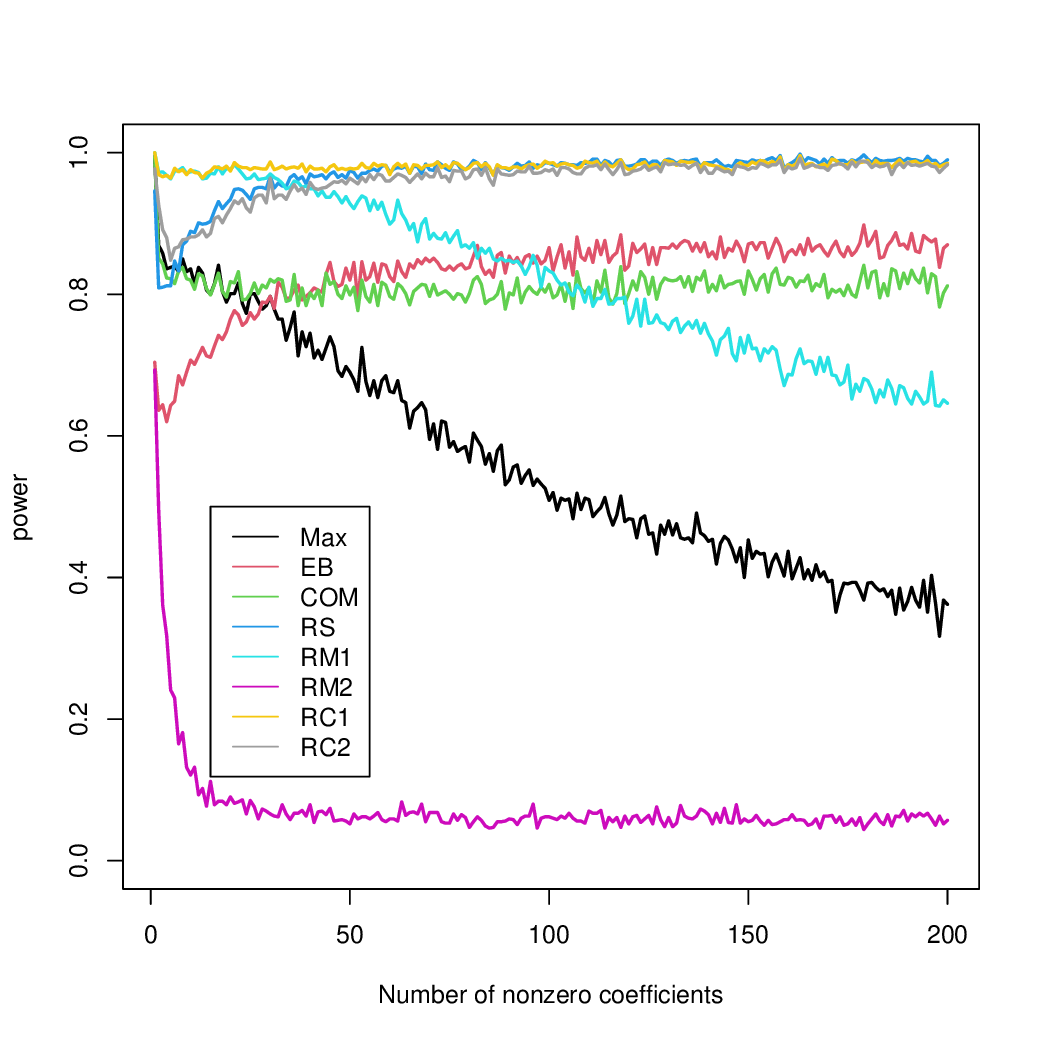}
  \caption{Mixture normal distribution}
  \label{fig:E4}
\end{subfigure}

\caption{Empirical power as a function of the number of nonzero coefficients of $\bm \beta$ under different error distributions.}
\label{fig:power_E1E4}
\end{figure}

Figure~\ref{fig:power_E1E4} reports the empirical power as a function of the number of nonzero coefficients $s$ under different error distributions. Across all four panels, classical sum-type procedures (e.g., EB/COM/RS) exhibit the expected advantage under dense alternatives but lose power noticeably in the sparse regime.
Conversely, max-type procedures (e.g., Max/RM1) are competitive when the signal is very sparse, yet their power deteriorates as the alternative becomes denser; moreover, the degradation is more pronounced under heavy-tailed errors in (E2)--(E4), reflecting their sensitivity to tail behavior. Notably, RM2 is uniformly underpowered in all panels, suggesting that the precision-adjusted max statistic is sensitive to the additional estimation step in these settings.
In sharp contrast, {RC1} delivers consistently strong power throughout the entire sparsity range in every panel, with only mild variation when moving from the Gaussian setting (E1) to heavy-tailed settings (E2)--(E4).
This uniform dominance indicates that {RC1 is simultaneously robust to the unknown sparsity level and to departures from normality}: it inherits sparsity adaptivity by effectively capturing both max- and sum-type evidence, while its rank-based construction stabilizes performance under heavy tails.
Overall, the results highlight {RC1 as the most reliable choice} when both the error distribution and the sparsity pattern of $\bm\beta$ are unknown.

To highlight the importance of precision-matrix transformation, we consider an additional configuration similar to \cite{XuZhou2021}. Specifically, we set
\[
\mathbf{\Sigma}=\big(0.5^{|i-j|}\big)_{1\le i,j\le p},  \bm\theta=\mathbf{\Sigma}\bm\beta,
\]
and define the signal vector as
\[
\bm\theta=2\sqrt{\frac{2\log(p)}{n}}\;(\theta_1,\ldots,\theta_p)^\top,
\]
where
\[
\theta_j=
\begin{cases}
1, & j\le m \text{ and } j \text{ is odd},\\
-1,& j\le m \text{ and } j \text{ is even},\\
0, & j>m.
\end{cases}
\]
All remaining settings are kept the same as in the previous experiment.

\begin{figure}[!htbp]
\centering
\begin{subfigure}[t]{0.48\textwidth}
  \centering
  \includegraphics[width=\linewidth]{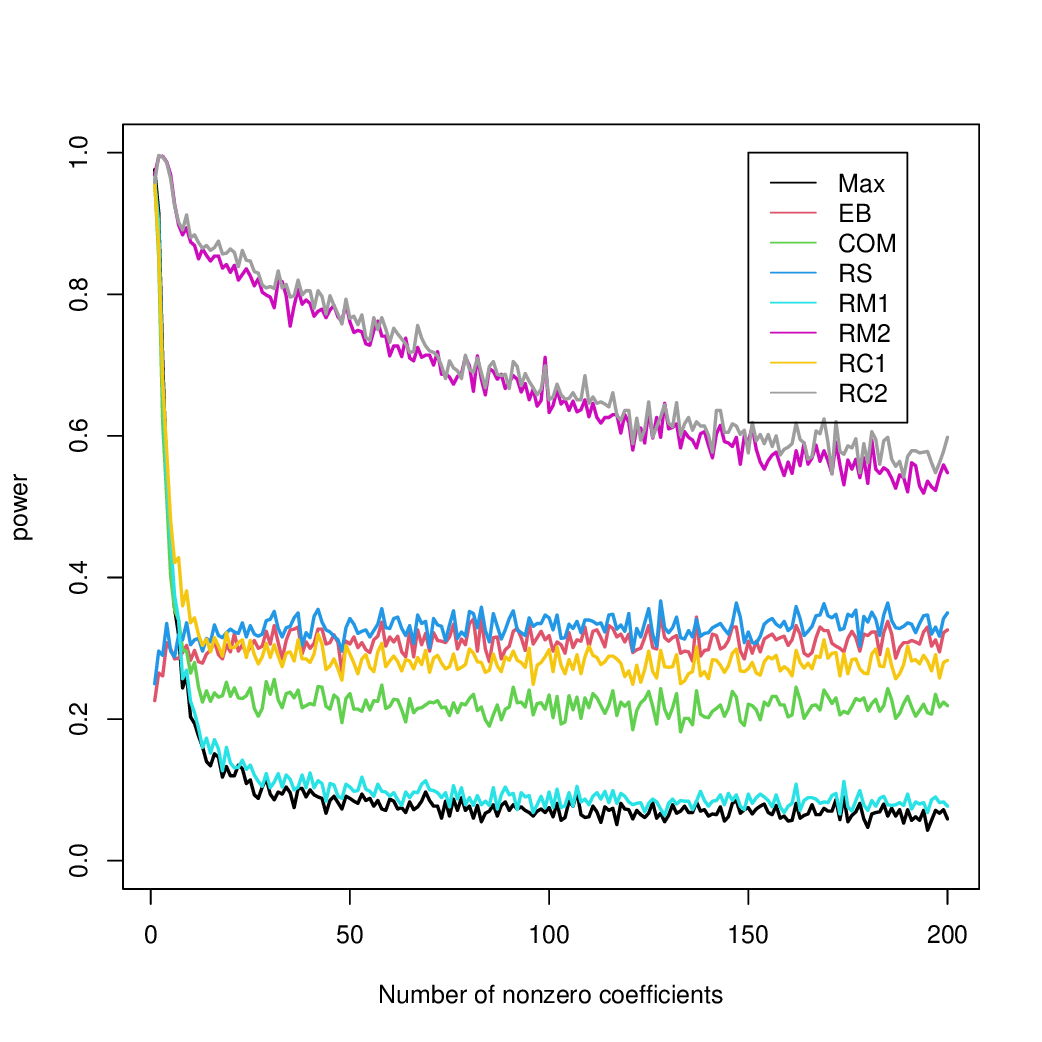}
  \caption{Normal distribution}
  \label{fig:E12}
\end{subfigure}\hfill
\begin{subfigure}[t]{0.48\textwidth}
  \centering
  \includegraphics[width=\linewidth]{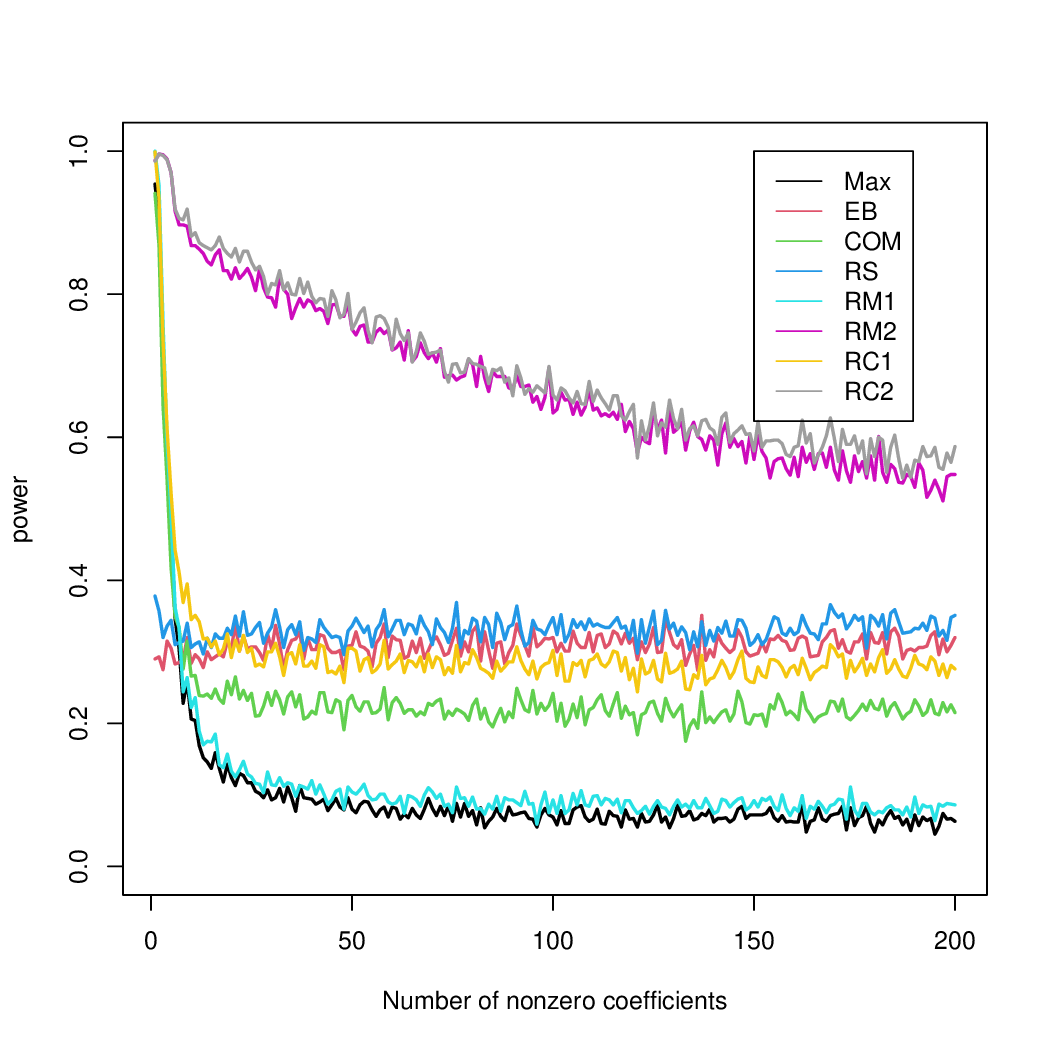}
  \caption{t(3) distribution}
  \label{fig:E22}
\end{subfigure}

\vspace{2mm}

\begin{subfigure}[t]{0.48\textwidth}
  \centering
  \includegraphics[width=\linewidth]{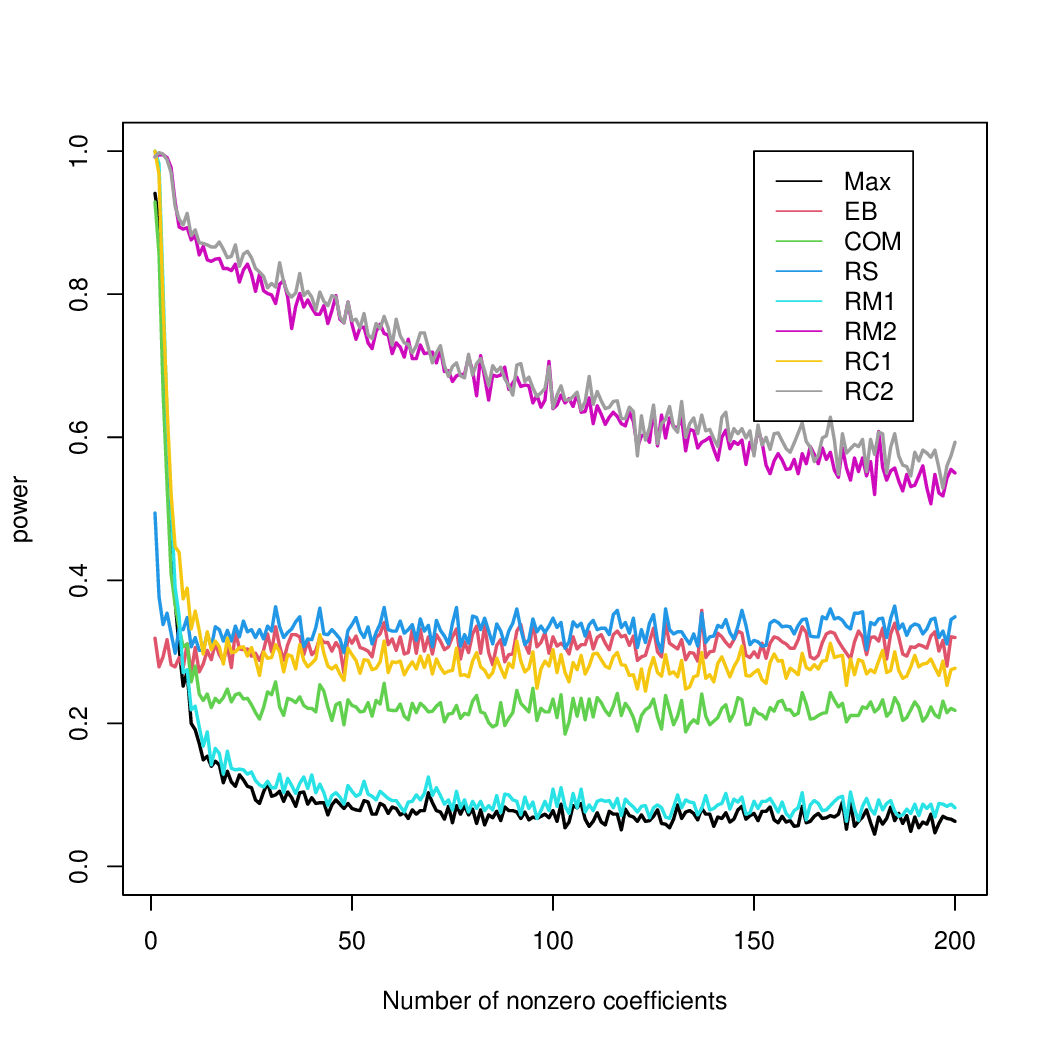}
  \caption{Lognormal distribution}
  \label{fig:E32}
\end{subfigure}\hfill
\begin{subfigure}[t]{0.48\textwidth}
  \centering
  \includegraphics[width=\linewidth]{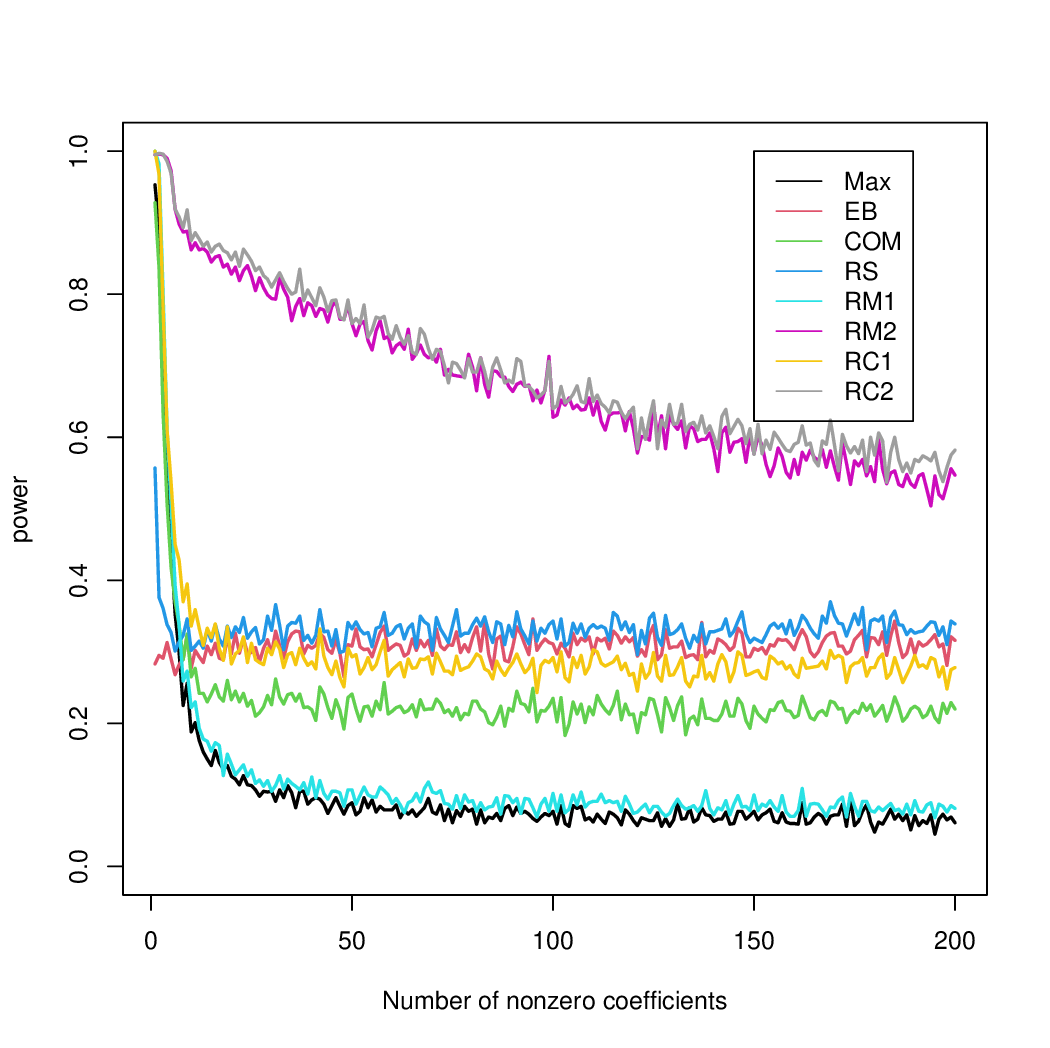}
  \caption{Mixture normal distribution}
  \label{fig:E42}
\end{subfigure}

\caption{Empirical power as a function of the number of nonzero coefficients of $\bm \theta$ under different error distributions.}
\label{fig:power_E1E42}
\end{figure}

Figure~\ref{fig:power_E1E42} reports the power curves for all competing procedures. Several clear patterns emerge. First, the precision-matrix--aware RM2 test is particularly effective under this Toeplitz-type dependence structure, delivering the best (or near-best) power across a wide range of sparsity levels. The Cauchy combination procedure RC2 exhibits a similar advantage, indicating that appropriately aggregating evidence after accounting for dependence can substantially improve detection performance. Second, when the errors are heavy-tailed, rank-based procedures remain more reliable than least-squares--based competitors. This robustness is most pronounced in the sparse regime (small $m$), where heavy-tailed noise can severely distort moment-based estimators and weaken the corresponding tests. In contrast, rank-based methods are less sensitive to tail behavior and thus maintain stronger power when signals are rare and the noise distribution is non-Gaussian.

\section{Conclusion}

This paper develops rank-based max-sum inference for testing high-dimensional regression coefficients under minimal moment conditions. By combining a sum-type rank statistic, which is sensitive to dense alternatives, with max-type rank statistics, which target sparse signals, we obtain procedures that adapt automatically to a wide range of sparsity regimes. We establish asymptotic null distributions for the proposed statistics under mild regularity conditions on the design and error distribution, allowing for heavy-tailed errors without requiring finite variance. A key theoretical contribution is an asymptotic independence result between the sum-type and max-type rank statistics, which enables simple and powerful Cauchy combination tests with asymptotically valid size. Numerical studies demonstrate that the proposed rank-based max-sum and Cauchy combination tests deliver competitive or superior power compared with least-squares based methods, especially under heavy-tailed or contaminated settings and when signals are sparse. Overall, the proposed framework provides a robust and practical toolkit for high-dimensional regression testing.
\section{Appendix}
\subsection{Proof of Theorem \ref{th1}}
Define $\bm \xi=\frac{1}{\sqrt{n}}\sum_{i=1}^n \bm X_i e_i=(\xi_1,\cdots,\xi_p)^\top$. Under the null hypothesis, we have $E(\bm \xi)=\bm 0$ and $\Cov(\bm \xi)=\mathbf{\Sigma}$. By Theorem 1 in \cite{ChenZhangZhong2010}, we have $\widehat{\tr(\mathbf{\Sigma}^2)}$ is a consistent estimator of ${\tr(\mathbf{\Sigma}^2)}$. So
\begin{align*}
T_{SR}=\frac{\sum_{i=1}^p \xi_i^2-\tr(\mathbf{\Sigma})}{\sqrt{2\tr(\mathbf{\Sigma}^2)}}+o_p(1).
\end{align*}
By Theorem 1 in \cite{XuZhou2021}, we have
\begin{align*}
T_{RM1}=\max_{1\le i\le p} \xi_i^2+o_p(1).
\end{align*}
Taking the same procedure as the proof of (7.15) in \cite{liu2024spatial}, we can show that
\begin{align*}
P\left(\frac{\sum_{i=1}^p \xi_i^2-\tr(\mathbf{\Sigma})}{\sqrt{2\tr(\mathbf{\Sigma}^2)}}\le x, \max_{1\le i\le p} \xi_i^2 \le y\right)\to P\left(\frac{\sum_{i=1}^p \zeta_i^2-\tr(\mathbf{\Sigma})}{\sqrt{2\tr(\mathbf{\Sigma}^2)}}\le x, \max_{1\le i\le p} \zeta_i^2 \le y\right)
\end{align*}
where $\bm \zeta=(\zeta_1,\cdots,\zeta_p)^\top \sim N(\bm 0,\mathbf{\Sigma})$. Finally, by Theorem 3 in \cite{FengJiangLiLiu2024}, we have
\begin{align*}
P\left(\frac{\sum_{i=1}^p \zeta_i^2-\tr(\mathbf{\Sigma})}{\sqrt{2\tr(\mathbf{\Sigma}^2)}}\le x, \max_{1\le i\le p} \zeta_i^2 \le y\right) \to \Phi(x)F(y).
\end{align*}
Here we complete the proof. \hfill$\Box$

\subsection{Proof of Theorem \ref{th2}}
By Theorem 2 in \cite{XuZhou2021}, we have
\begin{align*}
T_{RM2}=\max_{1\le i\le p} \nu_i^2/\omega_{ii}+o_p(1)
\end{align*}
where $\bm \nu=(\nu_1,\cdots,\nu_p)^\top =\mathbf{\Omega}\bm \xi$. And we can also rewrite $T_{SR}$ as 
\begin{align*}
T_{SR}=\frac{\bm \nu^\top \mathbf{\Sigma}^2\bm \nu -\tr(\mathbf{\Sigma})}{\sqrt{2\tr(\mathbf{\Sigma})}}+o_p(1).
\end{align*}
Taking the same procedure as the proof of (7.15) in \cite{liu2024spatial}, we can show that
\begin{align*}
P\left(\frac{\sum_{i=1}^p \bm \nu^\top \mathbf{\Sigma}^2\bm \nu-\tr(\mathbf{\Sigma})}{\sqrt{2\tr(\mathbf{\Sigma}^2)}}\le x, \max_{1\le i\le p} \nu_i^2/\omega_{ii} \le y\right)\to P\left(\frac{\bm v^\top \mathbf{\Sigma}^2\bm v-\tr(\mathbf{\Sigma})}{\sqrt{2\tr(\mathbf{\Sigma}^2)}}\le x, \max_{1\le i\le p} v_i^2/\omega_{ii} \le y\right)
\end{align*}
where $\bm v=(v_1,\cdots,v_p)^\top\sim N(\bm 0,\mathbf{\Omega})$.
Define the sum-type statistic
$$
S_p
=
\frac{\bm v^\top \mathbf\Sigma^2\bm v-\tr(\mathbf\Sigma)}{\sqrt{2\tr(\mathbf\Sigma^2)}},
$$
and the max-type statistic 
$$
M_p=\max_{1\le i\le p}\frac{v_i^2}{\omega_{ii}}.
$$
Let
$$
u_p(y)=2\log p-\log\log p+y,
\qquad
F(y)=\exp\Big\{-\pi^{-1/2}\exp(-y/2)\Big\}.
$$
Under Condition (C2), we have
$$
\tr(\mathbf\Sigma^2)\asymp p,
\qquad
\frac{\tr(\mathbf\Sigma^4)}{\tr^2(\mathbf\Sigma^2)}\to 0,
\qquad
\log p=o\big(\sqrt{\tr(\mathbf\Sigma^2)}\big).
$$
We next show that for any $x\in\mathbb R$ and $y\in\mathbb R$,
$$
P\Big(S_p\le x,\ M_p\le u_p(y)\Big)\to \Phi(x)F(y).
$$

Let $\bm Z\sim N(\bm 0,\mathbf I_p)$ and set $\bm v=\mathbf\Omega^{1/2}\bm Z$.
Then
$$
\bm v^\top\mathbf\Sigma^2\bm v
=
\bm Z^\top\mathbf\Omega^{1/2}\mathbf\Sigma^2\mathbf\Omega^{1/2}\bm Z
=
\bm Z^\top \mathbf\Sigma\,\bm Z,
$$
since $\mathbf\Omega^{1/2}\mathbf\Sigma^2\mathbf\Omega^{1/2}
=\mathbf\Omega^{1/2}\mathbf\Omega^{-2}\mathbf\Omega^{1/2}=\mathbf\Omega^{-1}=\mathbf\Sigma$.
Hence
$$
S_p=\frac{\bm Z^\top\mathbf\Sigma\,\bm Z-\tr(\mathbf\Sigma)}{\sqrt{2\tr(\mathbf\Sigma^2)}}.
$$
Let $\mathbf\Sigma=\mathbf U\mathbf\Lambda\mathbf U^\top$ with $\mathbf\Lambda=\mathrm{diag}(\lambda_1,\ldots,\lambda_p)$
and $\bm\eta=\mathbf U^\top\bm Z\sim N(\bm 0,\mathbf I_p)$. Then
$$
S_p=\frac{\sum_{j=1}^p\lambda_j(\eta_j^2-1)}{\sqrt{2\sum_{j=1}^p\lambda_j^2}}.
$$
By the Lyapunov (or Lindeberg) CLT for weighted sums of centered $\chi^2_1$ variables,
a sufficient condition is
$$
\frac{\sum_{j=1}^p\lambda_j^4}{\Big(\sum_{j=1}^p\lambda_j^2\Big)^2}
=
\frac{\tr(\mathbf\Sigma^4)}{\tr^2(\mathbf\Sigma^2)}\to 0,
$$
which holds under (C2). Therefore,
$$
S_p \cd N(0,1),
\qquad\text{i.e.,}\qquad
P(S_p\le x)\to \Phi(x).
$$
Write the standardized coordinates as
$$
\tilde v_i=\frac{v_i}{\sqrt{\omega_{ii}}},\qquad M_p=\max_{1\le i\le p}\tilde v_i^2.
$$
Under (C2), by Theorem 2 in \cite{FengJiangLiLiu2024},
$$
P(M_p\le u_p(y))\to F(y).
$$
Fix any finite index set $I\subset\{1,\ldots,p\}$ with $|I|=k$ fixed.
Let $\tilde{\bm v}_I=(\tilde v_i)_{i\in I}$.
We claim that for any bounded Lipschitz function $\varphi$,
$$
\sup_{\bm t\in\mathbb R^k:\ \|\bm t\|_\infty\le C\sqrt{\log p}}
\Big|
E\{\varphi(S_p)\mid \tilde{\bm v}_I=\bm t\}-E\{\varphi(S_p)\}
\Big|\to 0.
$$
To see this, use the regression decomposition for a standard Gaussian vector.
Let $\bm Z\sim N(\bm 0,\mathbf I_p)$ and note that each $\tilde v_i$ is a linear functional of $\bm Z$.
For a single index $i$, write $\tilde v_i=\bm q_i^\top\bm Z$ with $\|\bm q_i\|_2=1$.
Then
$$
\bm Z = t\bm q_i + \bm Z_\perp,
\qquad
\bm Z_\perp\sim N(\bm 0,\mathbf I_p-\bm q_i\bm q_i^\top),
\qquad
\bm Z_\perp\ \perp\ t,
$$
under the conditioning $\tilde v_i=t$.
Plugging into $\bm Z^\top\mathbf\Sigma\bm Z$ gives
$$
\bm Z^\top\mathbf\Sigma\bm Z
=
t^2\,\bm q_i^\top\mathbf\Sigma\bm q_i
+2t\,\bm q_i^\top\mathbf\Sigma\bm Z_\perp
+\bm Z_\perp^\top\mathbf\Sigma\bm Z_\perp.
$$
Hence the conditional mean shift satisfies
$$
\Big|
E(\bm Z^\top\mathbf\Sigma\bm Z\mid \tilde v_i=t)-\tr(\mathbf\Sigma)
\Big|
=
|t^2-1|\,\bm q_i^\top\mathbf\Sigma\bm q_i
\le C_1|t^2-1|,
$$
where $C_1=\sup_i \bm q_i^\top\mathbf\Sigma\bm q_i<\infty$ follows from (C2).
Therefore,
$$
\frac{\big|E(\bm Z^\top\mathbf\Sigma\bm Z\mid \tilde v_i=t)-\tr(\mathbf\Sigma)\big|}{\sqrt{2\tr(\mathbf\Sigma^2)}}
\le
\frac{C_1|t^2-1|}{\sqrt{\tr(\mathbf\Sigma^2)}}
=
o(1)
$$
uniformly over $|t|\le C\sqrt{\log p}$, since $\log p=o(\sqrt{\tr(\mathbf\Sigma^2)})$.

Similarly, one can show the conditional variance satisfies
$$
\frac{\Var(\bm Z^\top\mathbf\Sigma\bm Z\mid \tilde v_i=t)}{2\tr(\mathbf\Sigma^2)}=1+o(1)
$$
uniformly over $|t|\le C\sqrt{\log p}$, because the extra $t$-dependent terms contribute at most $O(t^2)$,
while $2\tr(\mathbf\Sigma^2)\asymp p$.
Moreover, $\bm Z_\perp^\top\mathbf\Sigma\bm Z_\perp$ is still a high-dimensional Gaussian quadratic form with the same
trace ratio condition $\tr(\mathbf\Sigma^4)/\tr^2(\mathbf\Sigma^2)\to 0$; thus the CLT remains valid conditionally.
By a standard smoothing argument (approximating indicators by Lipschitz functions), we obtain for each fixed $k$,
\begin{align}\label{st3}
\sup_{x\in\mathbb R}\sup_{\bm t:\ \|\bm t\|_\infty\le C\sqrt{\log p}}
\Big|
P(S_p\le x\mid \tilde{\bm v}_I=\bm t)-\Phi(x)
\Big|\to 0.
\end{align}
Define exceedance indicators
$$
A_i(y)=\mathbf 1\{\tilde v_i^2>u_p(y)\},
\qquad
N_p(y)=\sum_{i=1}^p A_i(y).
$$
Then $\{M_p\le u_p(y)\}=\{N_p(y)=0\}$.
The weak dependence in (C2) yields a Poisson limit
$$
N_p(y)\cd \mathrm{Poisson}(\lambda(y)),
\qquad
\lambda(y)=\pi^{-1/2}\exp(-y/2),
$$
hence $P(N_p(y)=0)\to e^{-\lambda(y)}=F(y)$.

To obtain the joint limit with $S_p$, similar to \cite{FengJiangLiLiu2024}, it suffices 
to show that for each fixed $k$ and distinct indices $i_1,\ldots,i_k$,
$$
P\Big(S_p\le x,\ A_{i_1}(y)=\cdots=A_{i_k}(y)=1\Big)
=
P(S_p\le x)\,
P\Big(A_{i_1}(y)=\cdots=A_{i_k}(y)=1\Big)
+o\Big(P(A_{i_1}(y)\cdots A_{i_k}(y))\Big).
$$
This follows from (\ref{st3}) by conditioning on $\tilde{\bm v}_I$ and noting that on the event
$A_{i_1}(y)=\cdots=A_{i_k}(y)=1$ we have
$\|\tilde{\bm v}_I\|_\infty=O_p(\sqrt{\log p})$.
Therefore the factorial moments of $N_p(y)$ computed jointly with $\mathbf 1\{S_p\le x\}$ factorize asymptotically, implying
$$
P\big(S_p\le x,\ N_p(y)=0\big)\to \Phi(x)\,e^{-\lambda(y)}=\Phi(x)F(y).
$$
Since $\{N_p(y)=0\}\equiv\{M_p\le u_p(y)\}$, the theorem follows.
\hfill$\Box$

\bibliographystyle{asa}

\bibliography{references}
\end{document}